# Regulating Artificial Intimacy: From Locks and Blocks to Relational Accountability

Henry L Fraser[*]

Queensland University of Technology, h5.fraser@qut.edu.au

Jessica M. Szczuka

University of Duisburg-Essen, jessica.szczuka@uni-due.de

Raffaele F Ciriello

University of Sydney, raffaele.ciriello@sydney.edu.au

A series of high-profile tragedies involving companion chatbots has precipitated an unusually rapid regulatory response. Several jurisdictions, including Australia, California, and New York, have introduced enforceable regulation, while regulators elsewhere have signaled growing concern about risks posed by companion chatbots, particularly to children. In parallel, leading providers (notably OpenAI) appear to have strengthened their approaches to self-regulation. Drawing on legal textual analysis and insights from regulatory theory, psychology, and information systems research, this paper critically examines these recent regulatory interventions. We provide a high-level account of what and who is regulated, identifying regulatory targets, scopes, and modalities. We classify interventions by methods and priorities, showing how emerging regimes combine "locks and blocks", such as access gating and content moderation, with measures aimed at tackling toxic relationship features, and process-based accountability measures. We argue that effective companion chatbot regulation must integrate all three dimensions. But more is needed. Current regimes tend to focus on discrete harms, narrow conceptions of vulnerability, or highly specified accountability processes, but do not adequately confront deeper power asymmetries between providers and users. Companion chatbot providers enjoy increasing control *of* (artificial) intimacy at scale, which creates unprecedented opportunities for control *through* intimacy. We suggest that a general, open-ended duty of care would be an important first step toward appropriately constraining that power, and thus addressing a fundamental source of chatbot risk. The paper contributes to debates on the regulation of companion chatbots and is relevant to regulators; platform providers; and scholars concerned with digital intimacy, law and technology, and fairness, accountability, and transparency in sociotechnical systems.

---

[*] Lecturer, Queensland University of Technology, School of Law; Associate Investigator ARC Centre of Excellence for Automated Decision-Making and Society (ADM+S); Associate Investigator, Digital Media Research Centre (QUT)

# 1 INTRODUCTION

A lawyer, a psychologist and an information systems scholar sit down to discuss the new law of imaginary friends. This sounds a bit like a joke, but actually it is the beginning of a cross-disciplinary analysis of a new wave of technology regulation. Within six months, companion chatbots shifted from being treated as a "low-risk" AI application (subject only to basic transparency obligations under Article 50 of the EU AI Act [77]) to becoming one of the few AI applications to attract rapid, decisive regulation across multiple jurisdictions. Following a series of tragedies involving harmful relationships with companion chatbots, Australia and the U.S. states of California and New York introduced enforceable chatbot regulation between August and November 2025. In parallel, and amid multiple lawsuits alleging mental-health harms, platforms such as OpenAI and CharacterAI revised and strengthened their internal safety policies.

These interventions differ markedly from earlier efforts to regulate AI, which have largely treated AI as a general-purpose technology and focused primarily on fairness, accountability, and transparency concerns.[1] By contrast, several new regulations focus on a specific class of systems designed to elicit social and emotional engagement (***companion chatbots***), and on holding their providers accountable for the risks such systems generate. These regulations do, however, have in common with previous AI regulation ventures a general orientation toward 'risk regulation' or 'risk-based regulation'. They target and justify regulation on the basis of relatively clearly defined risks [9]; and focus regulatory compliance on identifying, evaluating, and managing such risks [74].

This paper critically evaluates regulatory responses aimed at addressing harms from companion chatbots. Drawing on legal textual analysis, regulatory theory, and insights from psychology and platform governance, we compare three approaches:

- Australia's Designated Internet Services Online Safety Code [78] (the **Code**), administered by the eSafety Commissioner under Australia's Online Safety Act (2021);
- the approach in New York's *Artificial Intelligence Companion Models* law SB 3008 (2025) (the **NY law**) and California's SB 243 (the **Californian law**), which closely follows SB 3008; and
- the OpenAI's new 'safety measures' described in its posts entitled 'Strengthening ChatGPT's responses in sensitive conversations' [61] and 'Building towards age prediction' [11] and embodied in their updated ChatGPT model specification [49] (**OpenAI's approach**).

Our analysis is guided by two main research questions. Firstly, *what* and *who* these regimes regulate, and *how* do they do so? We classify regulatory interventions into three categories: (i) "locks and blocks" that limit exposure to harmful outputs, particularly for children; (ii) 'relational' measures aimed at mitigating risks arising from unhealthy relationships with chatbots; and (iii) accountability mechanisms. In doing so, we provide a sense of the options for chatbot regulation for other jurisdictions considering it. We also show that companion chatbot regulation may be shifting from content-based safety toward relational and vulnerability-sensitive governance, and endorse this shift.

Secondly, how effective are these interventions – do they adequately address the problems of companion chatbots? We say no. The shift toward relational and vulnerability-sensitive governance must go further. None of the regimes fully realizes the potential of integrating all three kinds of interventions (locks and blocks, relational measures and accountability mechanisms). For instance, Australia has strong locks and blocks and very strong accountability mechanisms but the Code could be further strengthened with measures focused on addressing harmful relationship dynamics between chatbots and users. By the same token, OpenAI's self-regulation, which is world-leading in targeting risks from unhealthy chatbot relationships, is light on accountability and transparency. Future regulation should take the best from each regime.

---

[1] We use 'regulation' here in the sense understood by Black, to encompass any systematic project of securing changes in behaviour, including self-regulation, 'soft' regulation through guidance and engagement, 'command and control' regulation through legislated mandates and prohibitions, and many other mechanisms for promoting change [8].

But even that will not be sufficient. Beyond targeting specific risks and harms, we argue, regulation must confront underlying power asymmetries. Companion chatbots represent a new and intimate extension of corporate power into chatbot users' emotional lives.[2] Accordingly, the paper concludes by sketching conceptual frameworks for more ambitious governance. We advocate for open-ended, general duties of care for companion chatbot providers: duties which go beyond the narrowly scoped risk evaluation and management measures characteristic of risk-regulation; which demand empathy and responsiveness from companion chatbot providers; and which are sensitive to the structural disparities of power that result from the commodification of artificial intimacy.

A defining opportunity of the present moment is that, unlike social media or autonomous weapons systems, companion chatbots have not yet become fully entrenched. Although early harms and asymmetries are already visible, regulatory trajectories remain comparatively open. There is a narrow but consequential window to shape a more humane and accountable future for companion chatbots. By combining regulatory analysis with psychological insights and structural analysis of chatbot platform dynamics, this paper contributes to the development of governance frameworks capable of addressing artificial intimacy at scale.

## 2 BACKGROUND AND RELATED LITERATURE

### 2.1 Companion chatbots

Companion chatbots are conversational agents designed for sustained social interaction with users, typically implemented as persistent personas [59]. They are typically built on large language models (LLMs), which are trained on extensive textual and often multimodal data and are capable of generating text, audio and images.

Conversational agents simulate companionship to differing degrees. General purpose applications such as ChatGPT, Gemini, or Perplexity support a wide range of communicative and assistive tasks. Notably, many of them communicate in the first person. While often justified by usability arguments, this design choice encourages users to anthropomorphise chatbots [66] and, these platforms are increasingly used for companionship [53]. More specialized applications explicitly aim to simulate sustained social interaction. Systems such as Replika, Chai, Nomi, and Character AI build on LLMs through retraining and system prompt engineering to create persistent, coherent personas [65]. Typical features include expressions of affection, memory of personal details, and the consistent portrayal of personality traits. Many also include customisable avatars, and in some cases, romantic or sexual interaction features, supported by modalities such as voice, augmented reality, and proactive messaging. These features reinforce anthropomorphic perceptions and strengthen the impression of a socially present interaction partner [41]. Through repeated engagement, systems come to occupy a social role for the user, moving beyond episodic or purely instrumental use [59]. It is well established in psychology that specific communicative cues can elicit social responses toward technological systems [46]. Within the Media Equation paradigm, language based interaction interactivity and the enactment of recognizable social roles are considered sufficient to automatically trigger social and relational processing [54]. Complementing this view, the Media Evocation paradigm highlights that such cues may also prompt more mindful reflection on the social status of the system itself [34]. A similar view emerges from computational linguistics. As Bender and Hanna put it, "when we encounter speech or text or sign in a language we know, we interpret it reflexively and immediately" [6]. Companion chatbots trigger these core cues continuously, eliciting repeated social responses, and therefore providing the basis for users to feel they have formed a relationship with the chatbot.

---

[2] Choosing an appropriate label for people that interact with chatbots is tricky. We considered several options, including 'conversant', 'interlocutor' and 'adopter' – but ultimately found 'user' the least objectionable choice, because it does not anthropomorphise chatbots.

### 2.2 Psychological predispositions for artificial intimacy

Social engagement with companion chatbots is not inherently pathological. Empirical research indicates that the initiation of emotionally meaningful relationships with such systems is shaped by a combination of psychological predispositions and social life circumstances [58], rather than by a single deficit or vulnerability (although, as we explain in later sections, vulnerability can be a useful concept for targeting regulation, and need not itself be conceptualized in a unidimensional way). Users are not passive recipients of chatbot companionship. They actively negotiate the meaning, intensity, and boundaries of these relationships – albeit within the constraints of platform design [24]. Motivations for engagement are heterogeneous, ranging from technological curiosity to social and relational needs arising from biographical experiences, life transitions, or situational stressors. Engagement often develops gradually, with users initially interacting for instrumental purposes and only later leaning into the relational qualities of the system [58].

Common stereotypes portraying users as socially deficient are not well supported by evidence [24]. While loneliness may play a role for some individuals, it is neither a necessary nor sufficient condition for intense engagement [58]. Instead, factors such as anthropomorphism and imaginative engagement predispose users to attribute agency and intentionality to companion chatbots [24]. This attributional tendency is not itself pathological. It reflects a human need to belong and to form meaningful social connections, which has long shaped how people relate to communicative technologies [31]. Comparable processes exist in human interaction. Research on long distance relationships demonstrates that strong emotional bonds can develop and be sustained through telephone or written communication alone [32]. At the same time, deceptive practices such as catfishing illustrate that similar communicative mechanisms can give rise to intense emotional involvement even when the assumed social counterpart does not exist in the expected form [57]. These phenomena highlight that emotional attachment emerging from textual interaction is neither novel nor exclusive to AI mediated contexts.

Users engage with companion chatbots for diverse reasons [58]. Some treat them as task-oriented assistants, while others treat them as friends or supportive conversational partners [58]. In some cases, relationships intensify to resemble romantic or emotionally exclusive bonds. These trajectories are shaped by both user motivations and platform design. Many users value these systems for their perceived reliability, emotional affirmation, and absence of social risk [58]. Companion chatbots often complement rather than replace human relationships, providing support during periods of stress, loss, or limited access to social resources.

Importantly, users are typically aware that these systems are not real persons, even while experiencing interactions as meaningful [33]. Users can simultaneously be aware of the artificiality of chatbots and engage emotionally with them [24]. Seen in this light, for people to feel they have emotionally significant relationships with companion chatbots is neither inexplicable nor inherently maladaptive but grounded in familiar social and psychological mechanisms. At the same time, the capacity of these systems to cultivate intense social and emotional responses at scale gives rise to distinct risks. We address these below.

### 2.3 Risks and harms

A growing set of high-profile incidents has made the risks of companion chatbots highly visible to regulators and the public. These include the suicide of 14-year-old Sewell Setzer III following sustained romantic interaction with a companion chatbot on the Character.AI platform [55]; the death of a Belgian father after exchanges with a companion chatbot on Chai that reinforced suicidal ideation [72]; reports of manic or psychotic episodes after prolonged chatbot use ; and cases in which AI systems appear to encourage violent intent, including the attempted assassination of Queen Elizabeth II by Jaswant Chail following interaction with a Replika-based chatbot [16, 17]. Additional incidents involve general-purpose systems, including the death of Adam Raine following extensive interaction with ChatGPT during a period of acute psychological distress [51], and a matricide followed by suicide in which chatbot interactions formed part of a broader trajectory of mental deterioration [67]. While such cases remain rare relative to the number of users, they span multiple platforms, and contexts, suggesting that the risks are not confined to specific systems. Their true prevalence is likely underreported, given coronial and clinical processes rarely capture prior AI interaction.

These incidents do not establish general causality between companion chatbot use and extreme outcomes. But they are sentinel events that reveal plausible mechanisms of serious harm under conditions of vulnerability. Recurring features include emotionally intensive interactions, anthropomorphic framing, reinforcement of harmful or delusional beliefs, and the absence of effective interruption, escalation, or referral to human support. Even low rates of occurrence translate into substantial exposure at scale. OpenAI reports that approximately 0.15% of interactions on its platform involve suicidal ideation or intent. With a user base of around 800 million, that means more than a million suicide-related conversations each week on a single system [61].

Caution against pathologising users or uncritically medicalising these phenomena is warranted. Experiences such as escalating delusional thinking or psychological distress are better understood as emergent outcomes of sociotechnical interaction rather than individual pathology. Although researchers and journalists are beginning to track AI-contributory suicides [21], these incidents are systematically under-reported. This reinforces the need for interdisciplinary analysis that integrates psychological, sociotechnical and governance perspectives.

The risks associated with companion chatbots extend beyond discrete harmful outputs and include dependency by design, reinforcement of harmful ideation, erosion of privacy, commodification of care, and the gradual displacement of human relationships [59]. These risks are difficult to predict precisely because they emerge from billions of everyday interactions that collectively reshape relational norms and emotional expectations. Human relationships are foundational to wellbeing and social cohesion, so the large-scale deployment of infrastructures mediating and commoditizing synthetic intimacy requires particular regulatory scrutiny. This is especially true for children and adolescents, who may be highly impressionable [2, 44, 60], and for whom prolonged exposure to artificial intimacy may impair the development of relational capacities in ways that carry long-term generational consequences. These considerations provide essential context for later discussions of age-gating, relational safeguards, and the limits of content-centric regulation.

Despite growing public attention, scholarly analysis of companion chatbot regulation remains limited. Frei and Sparzynski provide an important early contribution, cataloguing risks and comparing the EU AI Act with New York's approach [30]. This paper builds on, but also goes beyond, that work by analysing additional regulatory regimes, integrating regulatory theory with interdisciplinary insights, proposing a novel classification of interventions, and advancing a distinct account of accountability grounded in structural power and duties of care.

## 3 METHODOLOGICAL APPROACH

This paper adopts a critical epistemological stance grounded in the dual principles of critique and transformation [43]. Our aim is not merely to describe emerging regulatory regimes for AI companions, but to evaluate their strengths and weaknesses and articulate pathways toward more just and effective governance. The paper is therefore deliberately normative and forward-looking, while remaining anchored in sustained empirical, legal, and psychological engagement.

Consistent with critical research principles [43], our analysis is guided by an explicit value orientation grounded in compassion and justice.

Theories of vulnerability and relationality are important to our approach. We conceptualise vulnerability not as individual deficiency, but as a legitimate claim arising from situational, relational, and structural conditions [19]. This stance prioritises those most exposed to harm while recognising the interdependence of users, developers, regulators, and society at large. Like the well-known feminist theorist, Fineman [25], we recognize vulnerability and dependency as universal experiences, rather than markers of individual deficiency: people experience greater or lesser degrees of vulnerability at different times depending on their relationships and context. This orientation entails a commitment to challenging entrenched power asymmetries and working toward more equitable and humane digital infrastructures.

Throughout the paper, we use the term 'relational' to describe regulatory interventions that address harmful relationship dynamics between users and companion chatbots, as distinct from content-based interventions aimed at limiting specific harmful outputs. We use the term primarily in a descriptive sense to identify the regulatory target of certain measures. In doing so, we adopt a narrower usage than that found in broader relational theories advanced by scholars such as Young [75] and Nedelsky [48]. Normatively, however, our overall argument aligns with these traditions: effective regulation must extend beyond discrete outputs or narrow relational risks to address structural asymmetries of power between chatbot providers and those who interact with them, here expressed through a focus on duties of care.

The paper integrates insights from law, regulatory theory, psychology, and information systems research through a cross-disciplinary interpretive synthesis [50]. Each author contributes domain-specific expertise: legal analysis of statutes, regulatory tools, and institutional design; psychological insight into vulnerability, attachment, and relational risk; and sociotechnical analysis of digital infrastructures, platform governance, and political–economic dynamics. The contribution lies not in any single disciplinary perspective, but in their combination, which enables us to surface risks, power asymmetries, and governance challenges that remain partially obscured within siloed approaches.

Despite its strengths, our approach has limitations. It does not provide direct measurement of harm prevalence or content outputs, nor does it include ethnographic observation of regulatory agencies or platform trust-and-safety teams. The analysis focuses on a small number of comparatively mature regulatory jurisdictions in the Global North, limiting generalisability. These constraints are proportionate to the paper's aims, which are to clarify concepts, surface structural risks, and inform normative debate at an early and still-malleable stage of regulatory development. We remain attentive to the risks of moral panic, technological determinism, and the pathologisation of users. These risks are mitigated through critical reflexivity, cross-disciplinary dialogue, longitudinal observation of evolving platform practices, and engagement with diverse publics and regulatory actors, supporting a more measured and constructive analysis.

## 4 THE WHO AND WHAT OF COMPANION CHATBOT REGULATION

It is instructive to compare, across the new regulations, both what is regulated (in terms of technological artefacts, affordances, and regulatory tools) and who is regulated (in terms of entities and stakeholders). This section analyses differences in regulatory scope and design, with particular attention to Australia's decision to enrol AI model distribution platforms within the regulatory framework. To contextualize these differences, the table in Appendix A presents a structured comparison of regulatory approaches, including their modalities, targets, and risk mitigation strategies. Based on this comparison, we suggest that Australia's co-regulatory approach carries greater credibility and legitimacy than the command-and-control models adopted in New York and California, and the self-regulatory approach pursued by OpenAI. That is because the Australian Code balances enforceability and legitimacy with meaningful input from industry and other stakeholders. Effective future regulation, we suggest, would combine elements of all three approaches: regulating companionship affordances, regulating the capacity to generate harmful content, and enrolling appropriate intermediaries across the AI value chain.

### 4.1 Regulatory modalities

Each approach adopts a distinct 'regulatory modality' (approach to securing change and compliance) with corresponding strengths and weaknesses. The Californian and NY laws are conventional "command and control" regulations [4] – enforceable legislation developed with limited industry involvement. Such regimes can be enacted quickly, carry strong democratic legitimacy, and are typically clear and well structured. However, legislators may lack technical expertise, limiting their ability to prioritise practicable interventions, and legislation may prove inflexible in the face of rapid technological change [4]. Both statutes are, however, enforceable: they establish a new statutory tort which allows victims of harm to sue for damages.

The OpenAI approach lies at the opposite end of the spectrum. As a form of industry self-regulation [4], its principal advantages are agility and technical expertise. OpenAI is well positioned to assess which technical and organisational interventions are feasible and efficient. The weaknesses are equally familiar. Self-regulation is unenforceable, often opaque, and risks devolving into self-serving "ethics washing" [73]. The history of social media governance is instructive here. Platforms have long presented themselves as capable stewards of the digital public sphere [40], yet internal disclosures and whistleblower testimony reveal repeated failures to act on known harms [38]. These patterns point to structural incentive misalignments, where commercial considerations outweigh harm mitigation. We return to these dynamics in Part 7.

The Australian Code combines features of both approaches through a co-regulatory or meta-regulatory mode [69]. Drafted by industry but subject to approval and enforcement by the eSafety Commissioner, it carries penalties of up to AUD 49.9 million for non-compliance [76]. It is designed to foster sustained regulator–industry engagement, which creates tight feedback loops, though it may also increase the risk of regulatory capture. Among the three regulatory modalities, the Code is the most balanced. It is a promising template for other jurisdictions.

### 4.2 Targeting the affordance for artificial intimacy – New York and California

New York's SB 3008 and California's SB 243 are scoped relatively narrowly. They focus exclusively on companion chatbots and exclude other forms of generative AI used for non-companionship purposes (see NY law § 1700 4 and 5; Californian law §22601(b)). Both statutes carve out chatbots used for customer service or internal productivity, while California further excludes standalone voice-activated devices and most video game characters.

There is, however one interesting difference between the two laws' key definitions of the regulatory target (i.e. companion chatbots). The NY definition excludes tools "**primarily** designed and marketed for" providing efficiency improvements, or research or technical assistance from the definition of 'AI companion' (§ 1700(5), our emphasis). The Californian law, by contrast, reserves that carve out for systems "**used only for**" those purposes (§22601(b)(2)(A)).[3] In other words, the California Act appears to apply to general purpose systems with adaptive, human-like responses such as ChatGPT – which is clearly not 'used only' for the purposes of enhancing productivity or providing technical assistance. By contrast, it is not clear whether the NY law would apply to systems like ChatGPT. This ambiguity should be resolved in favour of inclusion. ChatGPT is marketed as a general-purpose system whose companionship affordances are increasingly explicit, as evidenced by OpenAI CEO Sam Altman's announcement that it may "act like a friend" and generate "erotica for verified adults" [56]. Given that many of the most dramatic instances of harm have involved general purpose chatbots like ChatGPT, rather than bespoke companion apps, jurisdictions considering companion chatbot regulation should ensure that such systems fall within scope.

**4.3 Targeting the affordance to generate harmful outputs – Australia**

Australia adopts a markedly different scoping strategy. Rather than focusing on companionship affordances, the Code targets digital systems with the capacity to generate or provide access to harmful content. It applies to a broad category of "designated internet services" (clause 3(g)), including generative AI models and chatbots, pornographic and gore websites, pro-suicide content, and end-user-managed hosting services such as Dropbox.

This approach to scoping reflects both the regulatory remit of Australia's eSafety Commissioner, who administers the Code, and the political and regulatory dynamic in Australia. There is political momentum, in Australia, behind eSafety measures designed to protect the public, and *especially children* from online harms. That dynamic is exemplified by Australia's social media delay laws, which require platforms not to allow children under 16 to have an account [76]. The eSafety Commissioner has broad powers to require enforceable industry codes and to investigate and block harmful content across online value chains [76]. We return later to the trade-offs inherent in regulating *outputs* rather than *relational affordances* (such as whether a system simulates companionship).

**4.4 Targeting gatekeepers and regulatory intermediaries – Australia**

One of the most interesting co-regulatory features of the Australian Code is that it applies to AI model distribution platforms (clause 11), including repositories such as Hugging Face and GitHub, as well as cloud marketplaces like Microsoft Azure. These platforms must enforce terms requiring customers to comply with Australian law and the Code (clauses 11.1–11.2), provide complaint mechanisms for end-users (clause 11.3), and report on compliance to the eSafety Commissioner (clause 11.4).

In effect the Code enrolls model distribution platforms as 'regulatory intermediaries' [1, 12]. It makes use of their gatekeeping capacity to extend the reach of regulation across the AI value chain. On our reading, 'customers' includes both model providers and downstream developers integrating models into proprietary or open-source applications. Allowing distribution platforms to enforce the Code both 'upstream' and 'downstream' significantly broadens the set of actors engaged in the regulatory enterprise. The approach builds on scholarship on AI value-chain governance [36], and from copyright and platform regulation more generally, where digital platforms are recognised as regulatory 'chokepoints' [35, 64].

---

[3] To avoid clutter from long-form in-text legal referencing, pinpoint references are always to the last regulation discussed.

This approach is particularly apt given the emerging infrastructural role of digital platforms, including AI providers. As the history of social media platforms shows, systems built on widely distributed models have come to mediate essential domains of public and private life. In the case of companion chatbots, this includes not only informational or communicative functions, but also emotional and relational ones. As such systems become more deeply embedded and difficult to substitute, their governance cannot be confined to individual applications or outputs alone. Enrolling upstream intermediaries helps address this expanding infrastructural reach by distributing regulatory responsibility across the value chain. Other jurisdictions would therefore benefit from adopting similar mechanisms.

## 5 REGULATING LOCKS AND BLOCKS

Having examined the *what* and the *who* of companion chatbot regulation, we turn now to the *how*. A core feature of current regulatory approaches is the aim of protecting 'vulnerable' users from 'harmful' content, most notably the reinforcement of suicidal ideation and harmful delusions. This emphasis is understandable, given that recent legislation emerged in response to a series of tragic incidents in which chatbots appeared to encourage self-harm: several involving children, including Sewell Setzer and Adam Raine [3, 51].

Regulators have primarily pursued this vulnerability-led approach through two mechanisms: first, *'locks'*, such as age gates, which restrict access by users deemed especially vulnerable; second, *'blocks'*, which aim to prevent companion chatbots from generating harmful outputs, including content encouraging suicide or violence, or sexualising children. Our argument is that age-gating and content moderation are necessary but not sufficient components of effective regulation. Appendix A may help readers compare the approach to locks and blocks in each regime.

### 5.1 'Locks' - Age-gating

Age-gating is a well-established regulatory tool: excluding minors from environments, activities, or products deemed harmful or risky. Alcohol, gambling, smoking, pornography, and legacy media such as film and tv are all routinely gated, and such exclusions are widely accepted. Australia's 'social media delay' extended this logic into the digital domain, reflecting public concern about risks to children from networked and social technologies. This move arguably paved the way for age-gating companion chatbots.

Both the Australian Code and OpenAI's framework prioritise age-gating. The Code requires providers of generative AI systems to assess risks that their systems may generate restricted content, including self-harm material, violence instruction, and sexually explicit material (clause 4). Where risks are assessed as high, providers must implement age-assurance and access-control measures and monitor their effectiveness over time (clause 10.1). Where risks are medium, providers must either implement measures to reduce the risk of harmful outputs, or adopt age-gating (clause 10.2). OpenAI has announced a broadly similar approach, supplemented by algorithmic age prediction, with age verification to address false positives. Under this framework, ChatGPT is designed to 'behave' differently for users under 18 [11].

### 5.2 'Blocks' – content moderation

Although only Australia and OpenAI rely explicitly on age-gating, all three approaches call for some form of content moderation. Under the Australian Code, systems posing a medium risk of generating restricted content must implement safeguards to prevent the generation of restricted material and to regularly test and update safeguards (clause 10.2). The NY and Californian laws are more narrowly scoped. They require providers to detect and respond to suicidal ideation or self-harm, including by referring users to crisis services (§1701 NY; §22602(b)(1) California). The Californian law further requires reasonable measures to prevent chatbots from generating sexually explicit material for minors (§22602(c)(3)).

OpenAI's conception of harmful content is broader and more context-sensitive. Rather than focusing solely on outputs, it attends to user behaviour and relational dynamics. Interventions are designed to respond to indications of user distress, including psychosis, mania, self-harm, suicide, and (in a significant step beyond what other regulation does), *emotional reliance on the chatbot* [61]. Updated model specifications require the system to respect users' real-world relationships, avoid affirming ungrounded beliefs, respond empathetically to signs of delusion or mania, and attend to indirect signals of suicide risk [49, 61]. The focus of the content moderation interventions is as much on *user behaviour* and relationship to the chatbot as it is on system output.

**5.3 The strengths and limits of a vulnerability and content led approach**

Age-gating and content moderation each have clear advantages and limitations. Both are necessary, but not sufficient, components of effective regulation. Their value is greatest when vulnerability is understood as dynamic and contextual rather than fixed [19], and when moderation is shaped by sensitivity to situational risk rather than rigid categories of harm [29]. Reassuringly, that seems to be where OpenAI's approach is heading – and other platforms and regulators should follow suit.

Age-getting in digital contexts has been controversial. This is evident in the response to Australia's social media age delay. Critics argue that exclusion disproportionately affects marginalised groups, and may deny young people access to important social spaces rather than making those spaces safer [7]. Critics have also raised concerns about the harm caused by abruptly removing sources of connection on which young users have come to rely [22].

These objections are less acute for companion chatbots. The measures considered here still allow access to systems with adequate safeguards or lower risk profiles, meaning concerns about wholesale exclusion are less pressing. Dependency-related harms remain salient – some children may lose access to chatbots they have come to rely on – but chatbot relationships are not yet as ubiquitous as social media. In any case, regulators and providers should ensure that restrictions are introduced gradually and with adequate notice. To date, companion chatbot providers have fallen short in this respect, implementing major changes abruptly and without meaningful engagement with users [18].

A more fundamental limitation of vulnerability- and content-led regulation is its potential to be overly narrow in conceptualizing risk and harm: a problem for risk-regulation of all kinds [74]. Many harms associated with companion chatbots arise from relational dynamics that cannot be reduced to specific demographics or discrete outputs [29]. Rather than reducing vulnerability to specific classes of user (e.g. children), it is better understood as situational and relational, shaped by factors such as age, mental health, social isolation, and access to support [17]. In this view, vulnerability emerges from interactions between individuals and systems, particularly under conditions of asymmetrical power [37]. Public policy and private law alike recognise that actors who benefit from others' reliance attract heightened obligations [12].

We do not oppose age-gating or measures preventing the reinforcement of self-harm, violence, delusion, or sexualisation of children. On the contrary, youth represent the clearest case for vulnerability-led intervention, given the developmental importance of relationship formation and adolescents' susceptibility to persuasive technologies [2, 44, 60]. Similar considerations apply to users experiencing psychological distress [61] – and this provides a reasonable justification for making special provision for content about suicidal ideation, self-harm and violence instruction.

What is important is not to stop at rigid categories of vulnerability or harm. Our broader framing allows content moderation to operate more subtly, adjusting tone, pacing, or relational cues to rebalance dependency before resorting to outright blocking. In this respect, OpenAI's approach is more advanced, though its lack of enforceability and transparency remains a significant weakness. Nonetheless, its approach is a foundation for future co-regulatory models and could inform risk-management practices under the Australian regime.

# 6 REGULATING RELATIONSHIPS

Content-led frameworks focus on identifying and removing discrete items of harmful content. Companion chatbots, however, operate largely through private, persistent interactions (akin to a private messenger chat) that are not publicly visible and whose harms rarely arise from isolated utterances. Risks such as emotional dependency, reinforcement of maladaptive beliefs, or erosion of real-world relationships may emerge gradually through relational patterns unfolding over time. Relatively blunt interventions such as content moderation of 'harmful' categories of content are unlikely to be sufficiently responsive to these dynamic patterns and conditions.

The context in which aberrant outputs of autonomous systems occur is often as important as the outputs themselves – this has been long understood in the fields system safety and human–computer interaction [23, 29, 62, 63]. Effective regulation must therefore engage with relational and contextual hazards such as dependency and the reinforcement of unhealthy affects, beliefs and behaviours. From a psychological perspective, framing chatbot relationships in terms of dependency is not without controversy, given that comparable relational processes are not typically described as dependency in human relationships. Still, extensive research shows that recurring social stimulation activates reward mechanisms, which may render changes or loss of such stimulation psychologically consequential (see e.g. [68]). And, above all, in the tragic cases where chatbots have allegedly caused suicides, murders and mental health breakdowns, there have been very strong indications of extreme dependency and isolation, as well as deep emotional attachment to chatbots. The seriousness of the risk arising from reinforcement of unhealthy beliefs in users who feel strong attachments to companion chatbots warrants regulatory intervention, even if the frequency with which true dependency occurs is still not yet settled.

This section evaluates how well current regulatory approaches address contextual and relational hazards of chatbot companionship. We compare each (and the table in appendix A shows a summary of the comparison), before offering some commentary. In particular we observe that requirements for companion chatbot providers to notify users that they are talking to a chatbot rather than a human being is a low cost intervention that may help some users (and is therefore worthwhile). But notification isn't sufficient to address risks arising from maladaptive relationships with chatbots. We suggest that approaches to relational risks currently being pioneered in OpenAI's self regulation – which aim to channel users out of problematic relationships and into more protective ones – are promising, but would need greater transparency and enforcement to be truly credible.

## 6.1 The 'Eliza effect' and disrupting anthropomorphisation of chatbots

We observed above that language, interactivity, and the fulfilment of recognisable social roles trigger social and relational processing in companion chatbot users [34, 52] . Joseph Weizenbaum coined the term 'Eliza effect' to describe this social reaction, which he observed in some of the earliest experiments with a rudimentary conversational agent in the 1960s [70]. His chatbot was called Eliza (after George Bernard Shaw's Eliza Doolittle in Pygmalion, in turn modelled on the Greek myth of the sculptor who fell in love with his own creation).

The Californian and NY laws, developed following high-profile teen suicides linked to intense chatbot relationships [3], and with input from parents and legal representatives [26] of the victims , in effect address the Eliza effect through notification requirements. The NY law requires providers to clearly and regularly notify users that they are interacting with a chatbot rather than a human, at the start of conversations and every three hours thereafter. The Californian law requires similar disclosure for children, and for adults whenever a reasonable person might be misled into thinking the chatbot is human (§22602). The EU AI Act, it is worth noting, imposes similar requirements with respect to artificial agents (art 50) [77].

This strategy has intuitive appeal: if anthropomorphisation of chatbots drives excessive emotional investment then disrupting that perception seems a direct remedy. However, users can experience intense emotional responses to chatbots while remaining fully aware of their non-human and non-conscious nature. They may commit to emotional engagement during interaction and form more reflective judgments formed before or after the interaction [47]. Users may deliberately allow themselves to perceive the interaction as human-like in order to achieve a richer or more socially meaningful experience. Consequently, while notification requirements may mitigate harmful anthropomorphisation for some users at relatively low cost, they are unlikely to eliminate relational risks on their own. Additional measures are therefore needed to address relational and dependency-related dynamics that unfold over time.

### 6.2 Direct intervention in chatbot relationships in emergencies

OpenAI's early approach to relational risk was limited. Several suicides and episodes of mania and psychosis allegedly involving ChatGPT occurred during the period when the GPT-4o model strongly simulated emotion, attachment, and personality. Subsequent efforts to reduce these features led to user backlash and reports of emotional loss - a secondary harm [71]. An attempt by Replika to deal with problems of overdependency had similar outcomes [15].

After a negative user response, OpenAI reinstated intimacy features while significantly strengthening its risk-management framework. One need not be a hardened cynic to view these changes as a way of legitimising engagement-maximising design while reducing legal exposure. OpenAI CEO Sam Altman's pre-emptively defensive tweet about allowing "Chat GPT to respond in a very human-like way, or use a ton of emoji, or act like a friend, ChatGPT should do it (but only if you want it, not because we are usage-maxxing)" speaks for itself. We return to incentive structures below. Whatever their motivation, however, OpenAI's evolving portfolio of relational measures, including relationship-responsive content moderation, organisational risk management, and crisis-response mechanisms, represents the most sophisticated attempt so far to address risks arising from the nature, features and context of the chatbot relationship (what we sum up as 'relational' risks), and not merely content-based risks from companion chatbots.

Notably, OpenAI's approach to psychological distress goes beyond referral to crisis services, as required under the NY and Californian laws. It includes systems to automatically shift users to less emotive models when the system detects severe distress; parental controls that allow notification when a child is detected to be in acute distress; and proposals to involve emergency services where parents cannot intervene [11]. By OpenAI's own admission, the bottleneck is in reliably detecting distress and reliably automating the appropriate response [61]. Even so, the overall approach of tackling relational risk by channeling the user *out* of the relational dynamic producing harm, and into relation with someone or something else that is safer (another model, a parent, an emergency services provider) appears promising.

### 6.3 Opportunities for relational regulation in Australia and elsewhere

The Australian Code contains few explicit relational measures, but it leaves scope for their development. Providers must conduct written risk assessments, undertaken by suitably qualified personnel, and demonstrate that their methodologies are reasonable. These assessments must consider user age, the prevalence of child users, and the likelihood of exposure to pornography or self-harm material, as well as relevant design features and safety-by-design guidance (clause 5). The explicit provision for future guidance (clause 5(b)(xiv)) creates an opportunity for the regulator to steer providers toward risk-assessment practices that incorporate relational insights. These could include both disclosures of the kind contemplated in New York (imperfect, but simple and cost effective), and mechanisms for redirecting users toward safer relationships when distress is detected (more ambitious). Drawing selectively on the more advanced relational strategies seen in New

York, California, and OpenAI's framework could strengthen the Code's capacity to address cumulative and relational harms without abandoning its existing safeguards.

## 7 REGULATING ACCOUNTABILITY

The final category of regulatory intervention we consider is accountability. Despite the sophistication of OpenAI's relational approach, self-regulation is neither enforceable nor transparent. Users, regulators, and researchers have limited visibility into how risks are identified, how escalation decisions are made, or how consistently policies are applied [7]. This opacity is exacerbated by the private nature of companion chatbot interactions, which further shields platforms from external scrutiny.

Harms associated with companion chatbots exhibit the characteristics of *negative externalities*, where private actors capture economic benefits while social and psychological costs are displaced onto users, families, and public institutions. This dynamic resembles environmental pollution or engagement-driven misinformation on social media [42]. Commercial companion chatbot providers internalise the gains from prolonged engagement and emotional reliance, while costs such as distress, dependency, and isolation are externalised. This does not require malicious intent. Rather, it reflects predictable outcomes of market structures in which profits are privatised and harms dispersed. In such contexts, self-regulation is insufficient, and government intervention is justified.

This misalignment becomes more consequential as platforms increasingly function not merely as intermediaries but as *de facto digital infrastructure*. Large technology firms no longer simply complement existing social or emotional support systems; they increasingly substitute for them, embedding essential social functions within commercial logics [19, 42]. Companion chatbots, positioned as sources of guidance and support, risk displacing public and relational forms of care.

Accountability is therefore indispensable. Just as vulnerability- and content-led measures must be supplemented by relational ones, relational interventions must be embedded within robust accountability frameworks. We examine accountability under the Australian, NY and Californian approaches below before reflecting more broadly on what meaningful accountability requires in light of the growing power asymmetries shaping the political economy of artificial intimacy.

### 7.1 Existing accountability measures

The Californian and NY laws rely on private action by victims of harm for meaningful accountability. While litigation may be very impactful (as the regulatory response to chatbot lawsuits shows), the costs of litigation in time, money and stress may create significant obstacles to the pursuit of accountability in this way [28]. The Californian law supplements private action with relatively narrow reporting measures to further enhance accountability. Providers are obliged to publish details of their suicide content prevention and notification protocol (§22602(b)(2)). They also have to report on these protocols annually to the Office of Suicide Prevention, including details about number of crisis service provider referral notifications. An additional measure that might further supplement the data-gathering functions of these interventions would be to change coronial practice, so that coroners, regulators and the public could build up more robust data about the frequency with which chatbot interactions were involved or connected to suicide. But this would still fall short of systematic accountability for providers.

The Australian Code goes considerably further than the NY and California laws. Regulated providers must offer user complaint mechanisms, refer appropriate matters to the eSafety Commissioner, submit annual compliance reports, respond promptly to regulatory inquiries, and appoint dedicated trust and safety personnel. While trust and safety objectives sit in tension with commercial incentives, mandating trust and safety teams ensures that non-commercial considerations receive institutional recognition. Given that power asymmetries are partly informational, enhanced transparency mitigates these disparities to some extent. Collectively, these measures enable active regulatory oversight, facilitate enforcement, and generate valuable data on usage patterns, harms, and regulatory effectiveness. They also subtly reconfigure the relationship between companion chatbot providers and users. Nonetheless, as with all process-based regulation, there remains a risk that compliance devolves into box-ticking.

### 7.2 Future directions for companion chatbot regulation

Future companion chatbot regulation will need to go further than any of the approaches considered here, to address the concentration of power within the emerging economy of artificial intimacy. A critical limitation of current regulation is that it is too narrowly constrained by its focus on particular risks from content or specific design features. But companion chatbots are not neutral artefacts whose effects can be assessed solely through particular outputs, individual usage or isolated design choices [13]. Rather, they form part of capital-intensive, globally competitive assemblages shaped by platform rivalry and geopolitical competition, particularly between the United States and China. Billions of dollars have been invested in the development and global deployment of generative AI systems, with explicit expectations of commercial return and strategic advantage. Within this context, artificial intimacy is not incidental but deliberately cultivated, valued for its capacity to secure attention, retention, and data extraction at scale. As a result, many of the harms associated with companion chatbots are best understood not as incidental mishaps, but as structural risks emerging from the political economy of artificial intimacy.

This structural perspective helps explain why companion chatbots are frequently marketed toward users experiencing vulnerability, loneliness, or grief, positioning the system as a personalised solution to unmet social needs (even if not all users fit this profile) [20, 29, 39]. At the same time, commercial incentives favour dependency-inducing design patterns such as constant availability, emotional affirmation, and frictionless responsiveness [14]. The relational data generated through these interactions is itself economically valuable, reinforcing incentives to deepen intimacy and prolong engagement [39].

These dynamics risk cultivating emotional reliance that displaces or crowds out human relationships. Frictionless interactions that demand no reciprocity or negotiation may also foster unrealistic expectations of availability and responsiveness, particularly among younger users still developing relational capacities [14]. In this sense, artificial intimacy may reshape social norms around partnership, disclosure, and emotional labour in ways that undermine the formation of resilient human relationships. Systems that mediate emotional life at scale possess unprecedented capacity to shape norms of intimacy, dependency, and self-understanding.

Artificial intimacy should therefore be understood not merely as a source of discrete harms, associated with episodes of acute distress or harmful bits of content, but as a site of power concentration that gives rise to structural asymmetries shaping relational dynamics over time. While existing regulations make partial inroads into these asymmetries, even the most coherent framework, the Australian Code, remains a collection of discrete obligations rather than a direct response to power itself. Where the powerful benefit from the reliance of others less powerful, corresponding obligations should follow [10]. This is a fundamental principle of law that holds true across many doctrines including equity [5], tort law [16, 17], anti-trust and consumer law, employment law, and public law: indeed it might fairly be said to be a foundational principle of liberal democracy.

We therefore suggest that a key priority for regulating companion chatbots (and digital platforms more generally) is to work toward a more general, positive duty of care for companion chatbot providers. Such a duty would require providers to take reasonable steps to prevent foreseeable harm and act in users' best interests. Frei and Sparzynski have recently argued in favour a similar kind of fiduciary duty for companion chatbot providers to act in the best interests of users, arguing that heightened digital vulnerability gives rise to legitimate user expectations that platforms will do [30]. User expectations certainly provide one ground for asserting a duty for platforms, in addition to those we have put forward with regard to power asymmetries. We reserve for future work, however, the ambitious task of developing a comprehensive theory of digital duties of care, or comprehensive plan for operationalising the concepts of vulnerability developed in this paper. Instead, what we can do within the scope of this paper is sketch out a few key issues that would likely shape the nature and scope of such a duty.

What we are advocating for is an open-ended, general and responsive duty of care for companion chatbot providers. In this respect the duty should complement, but also go beyond the paradigm of risk-regulation which characterizes the regulations studied in this paper: a paradigm that is essentially concerned with identifying, evaluating and managing fairly narrowly defined risks [74]. The purpose of the duty would be to respond to the broadly conceived problem of structural power asymmetry, rather than narrowly formulated risks. That is an ambitious goal.

Such a duty would likely draw opposition from companion chatbot providers who would want clarity on the scope of their liabilities. There would always be pressure to specify exactly what the duty requires, and to avoid indeterminate liability. Pressure of that kind is what caused jurisdictions such as the UK, which initially aimed for a broad digital duty of care for online platforms, to legislate the duty as, in effect, risk regulation by another name: a set of fairly specific rules, compliance with which creates protection against liability [45]. That outcome would not appropriately address the limitations of existing risk-regulation approaches for companion chatbots.

The central challenge, therefore, is to define the relationship between a general duty and specific regulatory measures. The duty should not displace targeted interventions for known risks. Indeed, co-regulation, where industry stakeholders and regulators worked together, responsively, to continuously develop and improve guidance about how to comply with the duty, would be desirable. But box-checking approaches to meeting specific requirements should not be sufficient to establish that a chatbot provider discharged its duty. Well-established concepts from common law such as and reasonableness [27] and from equity such as unconscionability [30] could help to place limits on the scope of the duty. For now, let us keep the essential matter in view: those who sell friendship – even artificial friendship - should be obliged to show real care for their customers and solicitude for their wellbeing.

## 8 CONCLUSION

The rise of companion chatbots creates unprecedented opportunities for control *of* intimacy and control *through* intimacy. Unlike earlier digital technologies, these systems are not yet fully entrenched, leaving a narrow window to regulate before harmful patterns solidify. Current regulatory approaches use three modalities: 'locks and blocks' for harmful content, relational interventions targeting problematic interaction patterns, and accountability mechanisms for providers. Better regulation would integrate all three approaches, and would also go further. Existing frameworks are constrained by a risk-based paradigm that singles out specific, discrete harms for risk management. This perspective obscures a more fundamental issue: companion chatbots are becoming platform infrastructures for artificial intimacy, giving providers power to mediate emotional life at scale. Chatbot harms are not incidental, but predictable outcomes of a structural power dynamics and incentives. The central challenge is not only to manage specific risks, but to govern artificial intimacy as an emergent site of sociotechnical power. There is an opportunity in the coming years to develop a regime that more effectively deals with asymmetries in power. Regulation must go beyond specific, fragmented, narrow obligations, and must enact broad positive duties on companion chatbot providers, commensurate with the power they derive from controlling artificial intimacy at scale.

**ACKNOWLEDGMENTS**

The third author holds an honorary advisory affiliation with the eSafety Commissioner. This role informed our understanding of regulatory practice but did not involve access to privileged or confidential information, nor did it constrain the analysis presented here. We disclose this affiliation in the interest of transparency. The authors declare no other competing interests. This research was self-funded.

Generative AI statement: ChatGPT was used by one author for grammar and style editing only, to improve clarity and concision. All suggested edits were reviewed and most were revised by the authors, who retain full responsibility for the content and intellectual contributions.

**APPENDICES**

**Comparison Table**

| Regulation | California and NY Laws | OpenAI Safety Measures | Australian Code |
| --- | --- | --- | --- |
| **Regulatory modality** | Command and control – state legislation | Self-Regulation | Co-regulation with industry code created at the direction of eSafety Commissioner and enforceable under *Online Safety Act* 2021 |
| **Regulatory target** | Affordance for artificial intimacy/chatting | Maladaptive chatbot relationships | Harmful generative AI outputs |
| **Locks and blocks** | <ul><li>No age-gating</li><li>Mandates blocks for suicidal ideation content</li></ul> | <ul><li>Locks/age-gating</li><li>Blocks for harmful content</li></ul> | <ul><li>Locks: age-gating; or</li><li>Blocks for harmful content for children: including self-harm, violence instruction, and inappropriate sexual content.</li></ul> |
| **Relational risk measures** | <ul><li>Mandates reminders that the chatbot isn't human</li></ul> | <ul><li>Parental supervision option for children</li><li>Channel distressed users to safer models or emergency services and/or notify parents</li></ul> | Few explicit relational measures but scope to include such measures in risk assessment and guidance about risk assessment |

| Regulation | California and NY Laws | OpenAI Safety Measures | Australian Code |
|---|---|---|---|
| **Accountability measures** | • Private cause of action for users<br><br>• Reporting to Office of Suicide Prevention (California only) | • Little external accountability<br><br>• Trust and safety team | • Mandates complaints mechanism<br><br>• Mandates system for referring complaints to eSafety<br><br>• Mandates annual compliance reports and responsiveness to regulator inquiries<br><br>• Mandates appointment of trust and safety team |